\documentclass[a4paper]{jpconf}
\usepackage{graphicx}
\begin{document}
\title{Top Quark Studies at D0}

\author{Reinhild Yvonne Peters}

\address{University of Manchester, School of Physics and Astronomy,
  Oxford Road, Manchester M13 9PL, England; also at DESY, Hamburg, Germany }

\ead{reinhild.peters@cern.ch}

\begin{abstract}
Years after its discovery in 1995 by CDF and D0, the top
quark still undergoes intense investigations at the Tevatron. Using up
to the full Run~II data sample, new measurements of top quark
production and properties by the D0 Collaboration are presented. In particular, the
first observation of single top quark s-channel production, the
measurement of differential $t\bar{t}$ distributions, forward-backward
$t\bar{t}$ asymmetry, a new
measurement of the top quark mass, and a measurement of the top quark
charge are discussed. 
\end{abstract}

\section{Introduction}
The top quark, discovered in 1995 by the CDF and D0 Collaborations at the
Fermilab Tevatron collider, is the heaviest known elementary
particle~\cite{cdftopdiscovery,d0topdiscovery}. Due to its high mass and short lifetime, the top quark
provides a unique environment to study a bare quark. It is  believed to play a special
role in electroweak symmetry breaking and provide a window to physics
beyond the standard model (SM). 

Despite the currently running LHC $pp$ collider being a top quark
factory, the study of the top quark using Run~II Tevatron $p\bar{p}$
collision events provides complementary information. Tevatron Run~II
with  a collision energy of 1.96~TeV (starting 2001
and ending September 30th, 2011)
provided $\approx 10.5$~fb$^{-1}$ of integrated luminosity for
each of 
the D0 and CDF experiments. 

In this article, latest studies in single top quark production and top quark pair
production at the D0 experiment are presented. The discussed analyses
are performed in dileptonic and semileptonic final states, where
either both or one of the $W$ bosons, coming  from the decay of the
top quark, decay into a charged lepton and associated neutrino.

\section{Studies in Single Top Quark Production}
Top quarks can be produced in pairs via the strong interaction, or
singly via the electroweak interaction. The latter occurs via s-channel, t-channel
and Wt-channel production. The Wt-channel has a negligible cross section at the
Tevatron, and only became relevant and observable at the LHC~\cite{cmswt}. 
 In 2009, CDF and D0 reported first observation of single top
quark production~\cite{cdfsingletop,d0singletop}, where the s- and
t-channel were measured together. For the observation, CDF and D0 were using up to $3.2$~fb$^{-1}$ and
$2.3$~fb$^{-1}$ of data, respectively. The measurement of single top quark
production is very challenging, as the main background from $W$+jets
events looks very similar to the single top signature. Various
multivariate techniques have been employed to distinguish the signal
from the large background. 
The t-channel on its own has been first observed by D0 in
2011~\cite{d0tchannel} using 5.4~fb$^{-1}$ of data. 

Only recently, also the s-channel was observed by performing a
combination of CDF and D0 measurements. The analyses employed for
combination use up to
the full Run~II data sample. Semileptonic
events are considered in the analyses by both collaborations, with the
addition of an analysis by CDF, where 
events with a missing transverse energy plus jet signature are used, adding acceptance of events in which the lepton is not directly reconstructed. The
events are required to contain at least two jets, of which one or two
have to be identified as $b$-jet. A multivariate
discriminant is build to separate s-channel signal from
background. In this analysis, the $t$-channel single top production
cross section was
set to its SM value.  The combined analysis results in a cross section of
$\sigma_{s}=1.29^{+0.26}_{-0.24}$~pb, which deviates with more than  6.3 standard
deviations (SD) from zero~\cite{tevschannel}.

\section{Studies in Top Quark Pair Production} 
The production of top quark pairs at the Tevatron is dominated by
$q\bar{q}$ annihilation with a fraction of approximately 85\%, and
consists of  $\approx 15$\%
gluon gluon fusion. At the LHC, these fractions are roughly 
inverse. Besides the different cross section of $t\bar{t}$
production at the Tevatron
relative to the LHC, these different fractions of production mechanisms
are one of the main reasons why many analyses are
complementary at the Tevatron relative  to the LHC. 

\begin{figure}[h]                                                                                                                                                                                                                              
\begin{minipage}{18pc}                                                                                                                                                                                                                         
\includegraphics[width=18pc]{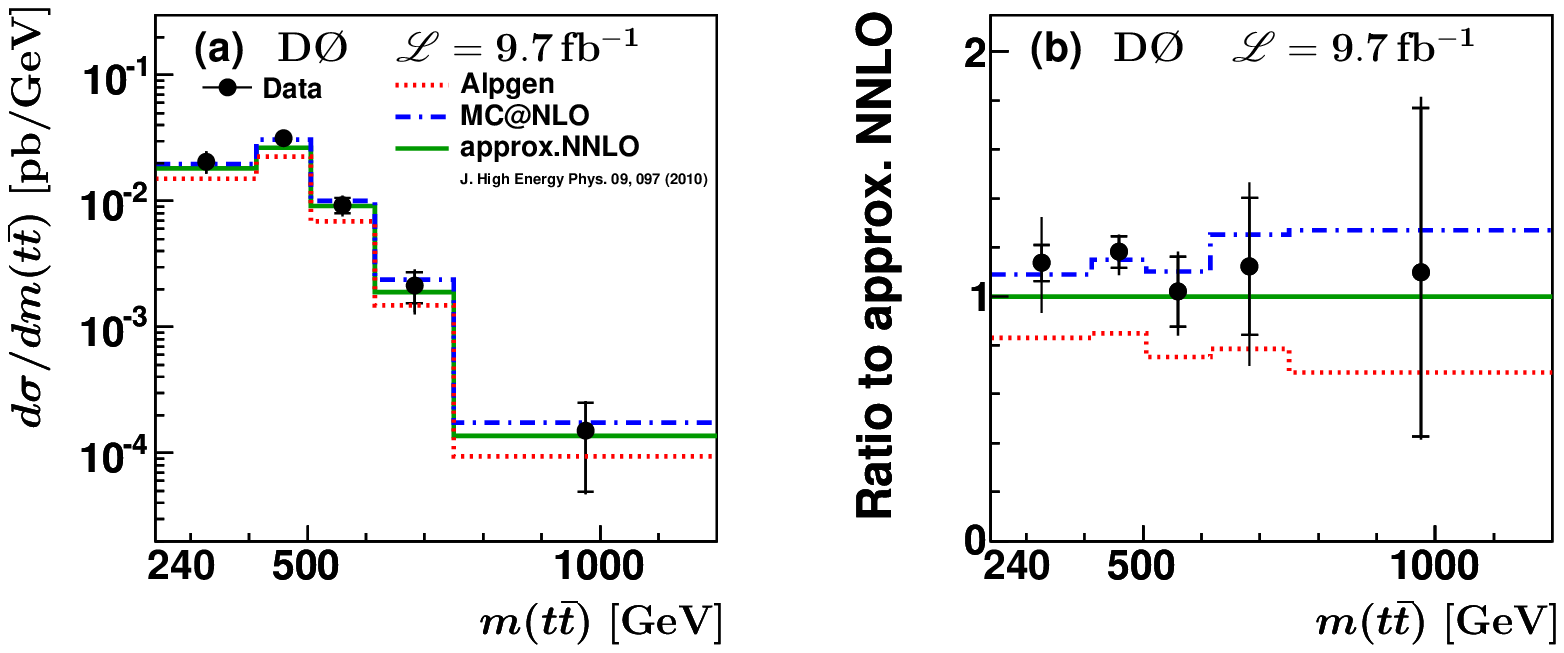}                                                                                                                                                                                                         
\caption{\label{mttbar} Differential $t\bar{t}$ distribution as
  function of $m_{t\bar{t}}$~\cite{diffxsec}.}                                                                                                                                                                          
\end{minipage}\hspace{2pc}%
\begin{minipage}{18pc}                                                                                                                                                                                                                         
\includegraphics[width=18pc]{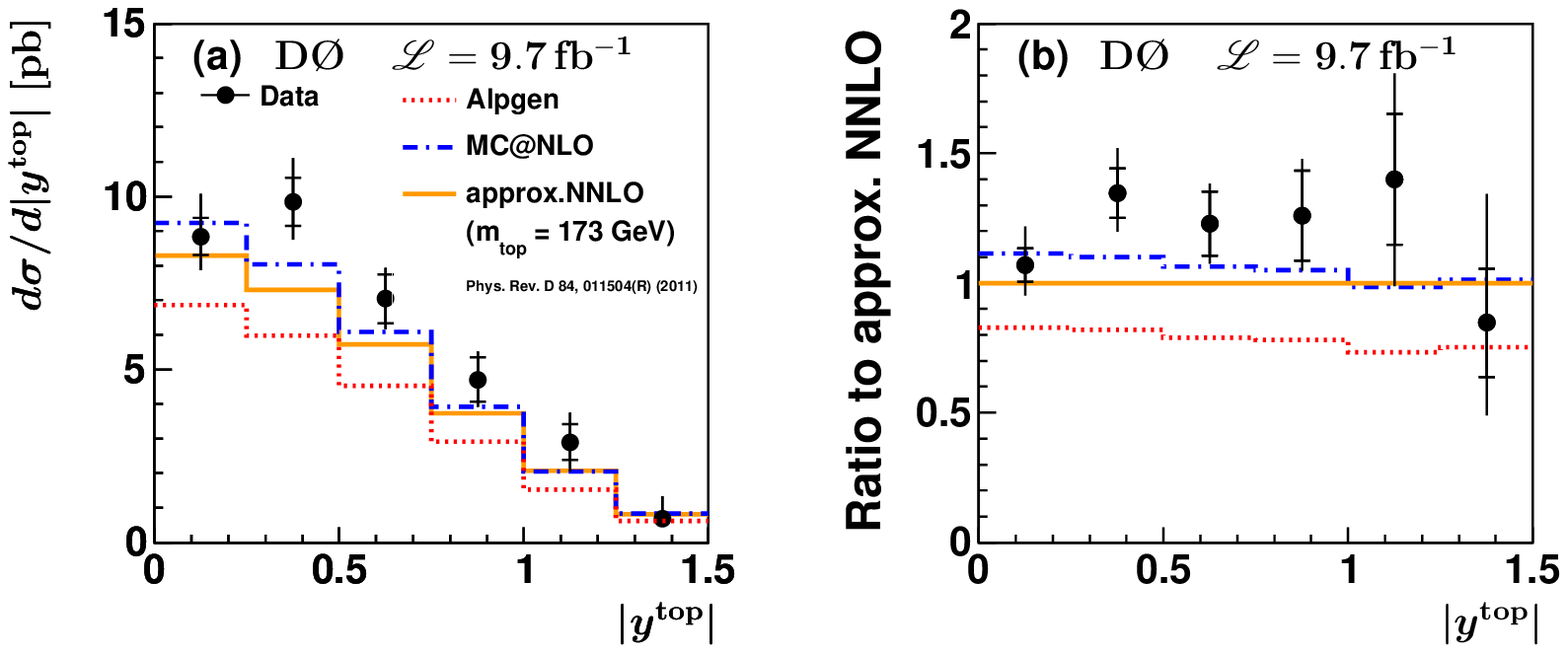}                                                                                                                                                                                                         
\caption{\label{deltay} Differential $t\bar{t}$ distribution as
  function of $|\Delta y^{top}|$~\cite{diffxsec}.}                                                                                                                                                                         
\end{minipage}                                                                                                                                                                                                                                                                                                                                                                                                                                               
\end{figure}

Recently, a new measurement of the $t\bar{t}$ cross section
inclusively and 
differentially as function of the invariant $t\bar{t}$ mass,
$m_{t\bar{t}}$, the rapidity of the top, $|y^{top}|$, and
 the transverse momentum of the top, $p_T^{top}$, has been performed by D0, using the full data sample of
  9.7~fb$^{-1}$~\cite{diffxsec}. The measurement of the $t\bar{t}$
  production provides a direct test of Quantum Chromo Dynamics (QCD) by
  confronting the measurement with SM predictions. The
  analysis uses lepton+jets events, where the
  $t\bar{t}$ event reconstruction is performed using a constrained
  kinematic fitter. The distributions are then corrected for detector
  and acceptance effects, using regularized unfolding, and are defined
  for parton-level top quarks including off-shell effects.   Using events with at least four jets,
  the inclusive cross section has been measured as $\sigma_{t\bar{t}}=
  8.3 \pm 0.7 {\rm (stat)} \pm 0.6 {\rm (syst)} \pm 0.5 {\rm
    (lumi)}$~pb, in good agreement with the SM prediction. Figure~\ref{mttbar} and Fig.~\ref{deltay} show the unfolded $t\bar{t}$
  distributions as function of $m_{t\bar{t}}$ and $|y^{top}|$,
  respectively, compared to approximate
  next-to-next-to-leading order (NNLO) calculations and different generator
  predictions. For these as well as the distribution as function of
  $p_T^{top}$, a good agreement between data and the NNLO calculations and
  generator predictions can be seen in general. In addition, a
  comparison to various axigluon models is performed. Using the
  differential distributions, these can be constrained. 

\begin{figure}[h]                                                                                                                                                                                                                              
\begin{minipage}{17pc}                                                                                                                                                                                                                         
\includegraphics[width=17pc]{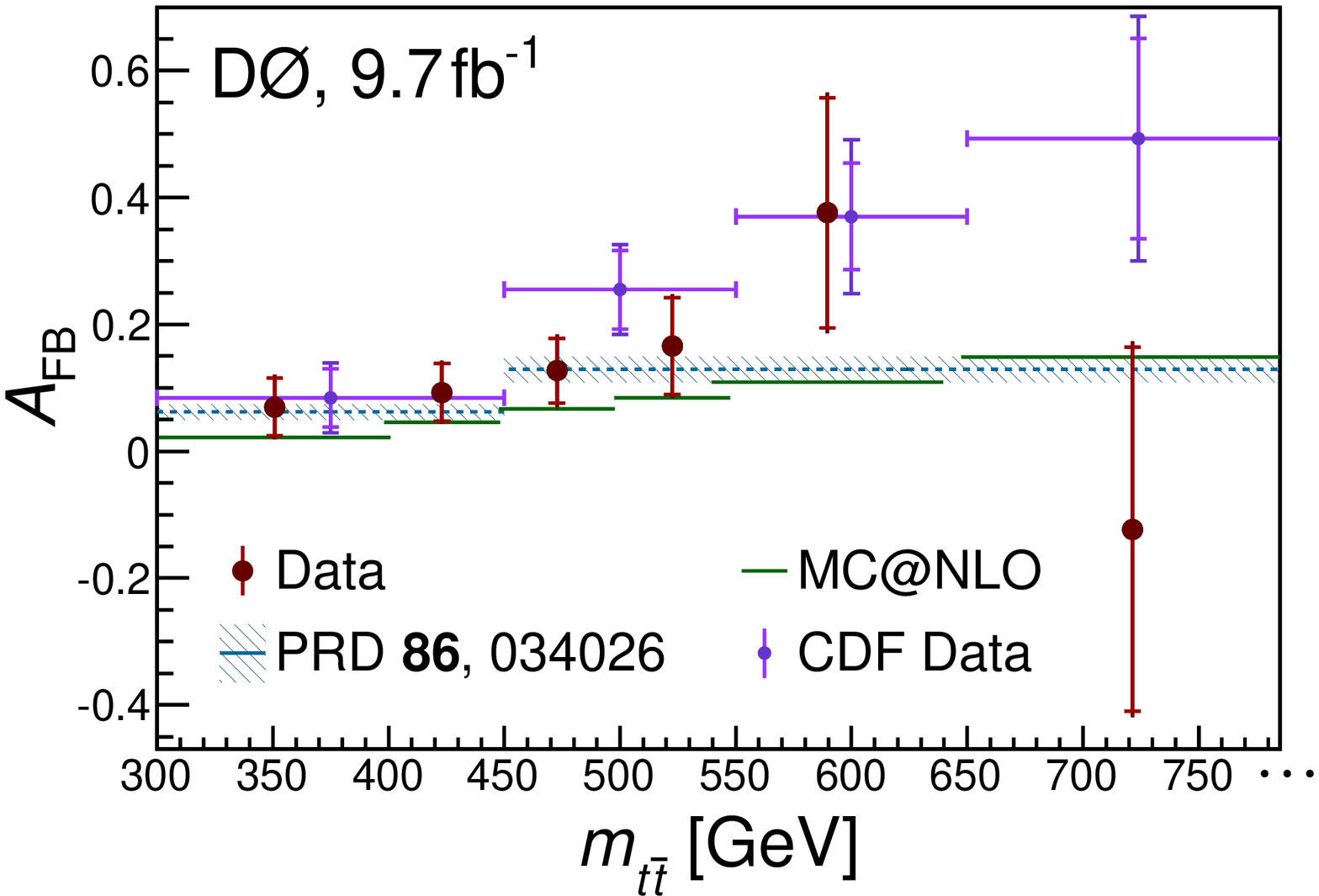}                                                                                                                                                                                                         
\caption{\label{asymmttbar} $A_{FB}^{t\bar{t}}$ as
  function of $m_{t\bar{t}}$~\cite{d0fbasym}.}                                                                                                                                                                          
\end{minipage}\hspace{2pc}%
\begin{minipage}{17pc}                                                                                                                                                                                                                         
\includegraphics[width=17pc]{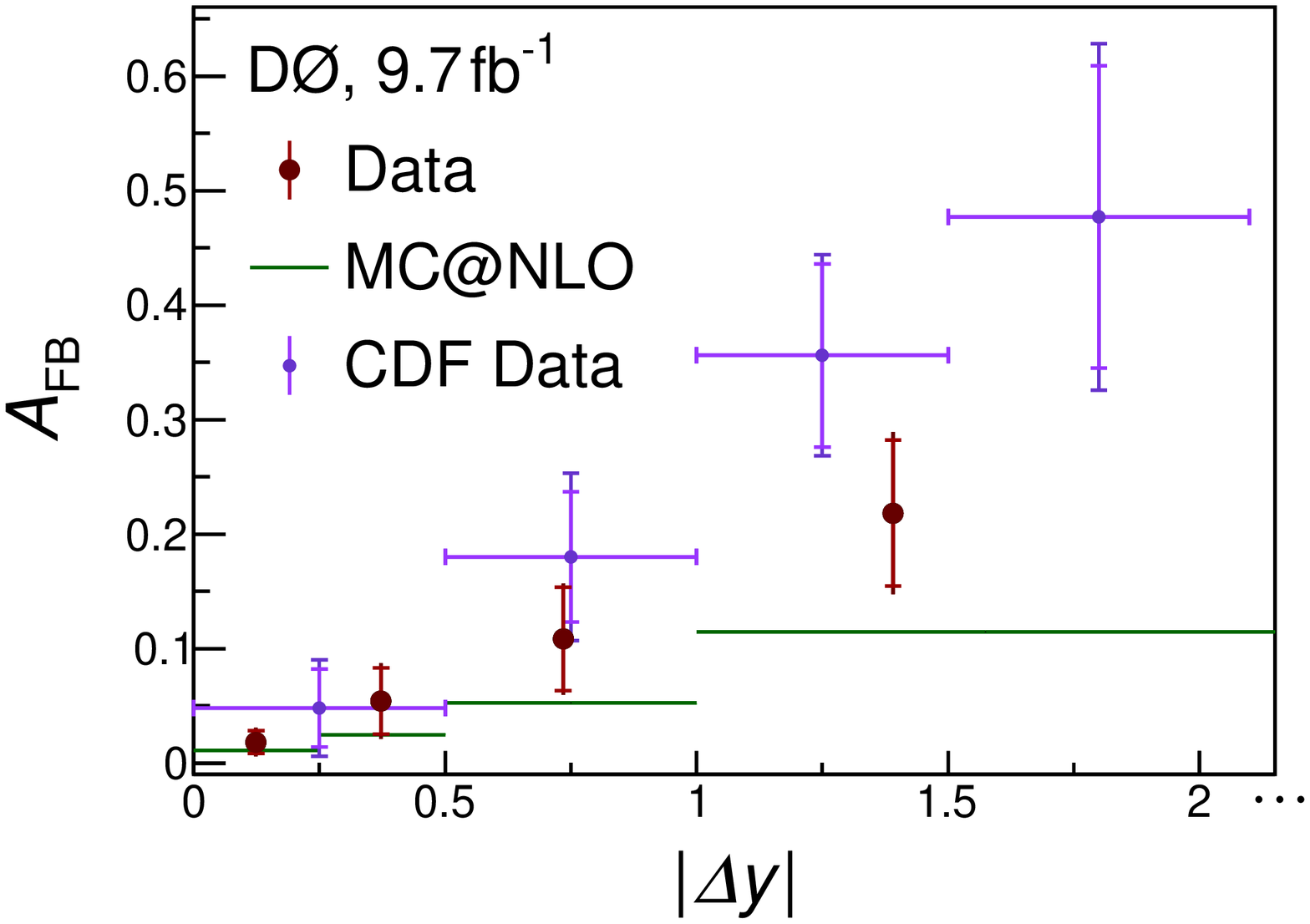}                                                                                                                                                                                                         
\caption{\label{asymdeltay} $A_{FB}^{t\bar{t}}$ as
  function of $|\Delta y^{top}|$~\cite{d0fbasym}.}                                                                                                                                                                         
\end{minipage}                                                                                                                                                                                                                                                                                                                                                                                                                                               
\end{figure}

An interesting feature of $t\bar{t}$ production is the
forward-backward asymmetry. At next-to-leading order (NLO), interference
between different $q\bar{q}$ diagrams causes a $t\bar{t}$ asymmetry, where the
top quarks are more likely to go into the direction of the incoming
quark. Various asymmetries can be studied, in particular the
forward-backward asymmetry
$A^{t\bar{t}}_{FB}= \frac {N(\Delta y >0) - N(\Delta y <0)} {N(\Delta
  y >0) + N(\Delta y <0)}$, with $N$ being the number of events with
the difference in top and antitop rapidity $\Delta y$ smaller or
larger zero, and the leptonic asymmetry $A^{l}_{FB}= \frac
{N(q_l y_l >0) - N(q_l y_l <0)} {N(q_l
  y_l >0) + N(q_l y_l <0)}$, with $q_l$ and $y_l$ being the charge and
rapidity of the lepton from $W$~boson decay, respectively. The
asymmetries have been measured using dileptonic and semileptonic
events, and get unfolded to production level. The combined leptonic asymmetry
in dileptonic and semileptonic events, using 9.7~fb$^{-1}$ of D0 data,
yields $A_{FB}^{l}= 4.7 \pm 2.3 {\rm (stat)}\pm 1.5 {\rm (syst)}$~\%~\cite{d0dilepasym,d0asym}, in
good agreement with the SM prediction at NLO QCD, including
electroweak (EW) corrections, of $A_{FB}^{l}= 3.8 \pm
0.2$~\%~\cite{bernreutherasym}. The asymmetry is also measured as
function of lepton $p_T$, showing good agreement with the prediction
from MC@NLO~\cite{mcnlo}.
The measurement of the forward-backward asymmetry in the
semileptonic channel explores events with three or at least four jets. The
measured asymmetry is $A_{FB}^{t\bar{t}} = 10.6 \pm 2.7 {\rm (stat)}
\pm 1.3 {\rm (syst)}$~\%~\cite{d0fbasym}, in good agreement with the NLO+EW SM prediction of 
 $A_{FB}^{t\bar{t}} = 8.8 \pm 0.6$~\%~\cite{bernreutherasym}. The
 measurement of $A_{FB}^{t\bar{t}}$  as function of $m_{t\bar{t}}$ and
   $|\Delta y|$ is shown in Fig.~\ref{asymmttbar} and
   Fig.~\ref{asymdeltay}, respectively. The differential
   $A_{FB}^{t\bar{t}}$ distributions as function of the two observables show good agreement with the prediction from
   MC@NLO, but are also consistent with the recent measurement by CDF~\cite{cdfasym}.

Another important analysis is the precise measurement of the top
quark mass. The top quark mass is a free parameter in the SM. Together
with the mass of the $W$~boson, the top quark mass constrains the
Higgs boson mass, and thus provides a self-consistency check. 
The D0 collaboration recently performed a new measurement of the top
quark mass, using  semileptonic events with at least one identified $b$-jet in the full Run~II data sample of 9.7~fb$^{-1}$. For
this, the matrix element technique is applied, which uses the full
event kinematic and thus provides the most precise method. In this
method, a probability is calculated for each event, where
a phase space integration over  leading order matrix elements, folded with parton
distribution functions and transfer functions is performed. For the
new analysis, the speed of the integration has been improved with
respect to the previous implementation in D0. Furthermore,  the
handling of the systematic uncertainties has also been improved,
allowing less conservative estimations of several sources of
systematic uncertainties. To reduce the uncertainty from jet energy
scale, jets from the hadronically decaying $W$ boson are used as an in-situ
constraint. The top quark mass is measured to be $m_t=174.98 \pm 0.58
{\rm (stat+JES)} \pm 0.49 {\rm (syst)}$~GeV~\cite{d0mtop}. With a
relative  uncertainty of 0.43~\%, this is the most precise single
measurement of the top quark mass to date. The main systematic
uncertainties originate from uncertainties on the residual jet energy scale, and
hadronization and underlying event. Using this new measurement as well
as updates on all-hadronic and dileptonic measurements by CDF, an
updated Tevatron top quark mass combination has been performed. Using
the BLUE method~\cite{blue} for the combination of Run~I and Run~II
results from CDF and D0 in various final states, the new Tevatron
combination yields  $m_t=174.34 \pm 0.37
{\rm (stat)} \pm 0.52 {\rm (syst)}$~GeV~\cite{tevmtop}. 

In the SM, the top quark has an electric charge of $+2/3$ of the electron
charge. An exotic model could be possible though, where the top quark
carries a charge of $-4/3$ the electron charge. Using 5.3~fb$^{-1}$ of
data, D0 performed a new measurement of the top quark charge in
semileptonic $t\bar{t}$ events. In this analysis, at least four jets,
of which at least two jets have to be identified as $b$~jets, are
required. To assign the final state objects to come from the top or
antitop quark, a kinematic fitting algorithm is used. The top quark charge
can then be determined by combining the charge of the lepton from the
$W$~boson decay with the charge of the $b$ jet. The determination of
the latter happens via a jet-charge algorithm, where a weighted
sum of the charge of the tracks belonging to the $b$~jet is
calculated. The weight used in the sum is $p_T^{0.5}$ of the
respective tracks. The factor $0.5$ in the exponent has been optimized
using $t\bar{t}$ events. The calibration of the jet charge algorithm is done using dijet events, where soft lepton and lifetime $b$-jet
identification was applied. A SM and exotic top charge template is then
constructed in $t\bar{t}$ events, and the fraction $f$ of the events with
SM top charge is extracted using a binned maximum likelihood fit. The
measured fraction $f$ yields $f=0.88 \pm 0.13 {\rm (stat)}\pm 0.11
{\rm (syst)}$~\cite{d0charge}. The dominant systematic uncertainty in this measurement
is the statistics of the dijet sample. This value of $f$ translates into the
exclusion of the hypothesis that top quarks carry a charge of $-4/3$
of the electron charge  at more than five SD, confirming earlier results by CDF and ATLAS. An alternative interpretation of the measurement can be
done, assuming that the $t\bar{t}$ sample is a mixture of top quarks
with charges $+2/3$ and $-4/3$. This measurement results in an upper
limit on the admixture of exotic top quarks of $f<0.46$ at 95\%
confidence level.

\section{Conclusion}
Despite the Tevatron having been switched off about three years ago,
the exploitation of the data are still ongoing. Many new
measurements of top quark production and properties have been released
recently by the D0 collaboration. Several of these are Tevatron
legacies, for example the most precise single measurement of the
top quark mass, the final word from D0 on the forward-backward
asymmetry, and the measurement of a variety of differential $t\bar{t}$
distributions. 

\ack
I would like to thank my collaborators from the  D0 collaboration for their help
in preparing the presentation and this article. I also
thank the staffs at Fermilab and collaborating institutions, and acknowledge the support from
the Helmholtz association.


\section*{References}
\begin{thebibliography}{9}
\bibitem{cdftopdiscovery} F.~Abe {\it et al.}  [CDF Collaboration], Phys.\ Rev.\ Lett.\  {\bf 74}, 2626 (1995).
\bibitem{d0topdiscovery}  S.~Abachi {\it et al.}  [D0 Collaboration],
  Phys.\ Rev.\ Lett.\  {\bf 74}, 2632 (1995).
\bibitem{cmswt} CMS Collaboration, Phys. Rev. Lett. {\bf 112},  231802 (2014).
\bibitem{cdfsingletop}  T. Aaltonen {\it et al.} [CDF  Collaboration], Phys. Rev. Lett. {\bf 103}, 092002 (2009).
\bibitem{d0singletop}  V. M. Abazov {\it et al.} [D0  Collaboration], Phys. Rev. Lett. {\bf 103}, 092001 (2009).
\bibitem{d0tchannel} V. M. Abazov et al. [D0  Collaboration],  Phys. Lett. B {\bf 705}, 313 (2011).
\bibitem{tevschannel} T. Aaltonen {\it et al.} [CDF and D0  Collaborations],  Phys. Rev. Lett. {\bf 112} , 231803 (2014). 
\bibitem{diffxsec} V. M. Abazov {\it et al.} [D0  Collaboration],  arXiv:1401.5785 [hep-ex] (submitted to PRD).
\bibitem{d0dilepasym} V. M. Abazov et al. [D0  Collaboration],  Phys. Rev. D {\bf 88}, 112002 (2013).
\bibitem{d0asym} V. M. Abazov {\it et al.} [D0  Collaboration],  arXiv:1403.1294 [hep-ex] (submitted to PRD). 
\bibitem{bernreutherasym}  W.~Bernreuther and Z.~-G.~Si, Phys.\ Rev.\ D {\bf 86}, 034026 (2012).
\bibitem{d0fbasym}  V. M. Abazov {\it et al.} [D0  Collaboration],  arXiv:1405.0421 [hep-ex] (submitted to PRD). 
\bibitem{mcnlo}  S.~Frixione and B.~R.~Webber,  J. High Energy Phys. {\bf 06}, 029 (2002).
\bibitem{cdfasym} T. Aaltonen  {\it et al.} [CDF  Collaboration], Phys. Rev. D {\bf 87}, 092002 (2013).
\bibitem{d0mtop} V. M. Abazov et al. [D0  Collaboration], Phys. Rev. Lett. {\bf 113}, 032002 (2014).
\bibitem{blue} L. Lyons, D. Gibaut, and P. Clifford,
  Nucl. Instrum. Meth. A {\bf 270}, 110 (1988); A. Valassi,
  Nucl. Instrum. Meth. A {\bf 500}, 391 (2003).
\bibitem{tevmtop} T. Aaltonen {\it et al.} [CDF and D0  Collaborations],  arXiv:1407.2682 [hep-ex].
\bibitem{d0charge} V. M. Abazov et al. [D0  Collaboration],   arXiv:1407.4837 [hep-ex] (submitted to PRD RC). 
\end{thebibliography}
\end{document}